\documentclass[aps, twocolumn, prd, preprintnumbers, amsmath, amssymb, amsfonts, superscriptaddress, nofootinbib]{revtex4-2}
\usepackage{amsmath,amssymb,bm,mathrsfs}
\usepackage{hyperref}
\usepackage{url}
\usepackage{graphicx}
\usepackage{color}
\usepackage{tikz}
\usepackage[compat=1.1.0]{tikz-feynman}
\usepackage[normalem]{ulem}

\begin{document}
\hfill TU-1299, KEK-QUP-2026-0004

\title{
Naturally Light Distortion
}

\author{
Kazunori Nakayama
}
%\email{kazunori.nakayama.d3@tohoku.ac.jp}
\affiliation{Department of Physics, Tohoku University, 
Sendai, Miyagi 980-8578, Japan} 
\affiliation{International Center for Quantum-field Measurement Systems for Studies of the Universe and Particles (QUP), KEK, Tsukuba, Ibaraki 305-0801, Japan}

\begin{abstract}

In the most general formulation of gravity, the metric and connection are independent degrees of freedom, and the connection may include torsion and non-metricity (or distortion, collectively) degrees of freedom, resulting in a huge number of possible dynamical fields. However, most fields are either non-dynamical or extremely heavy and the general relativity is recovered at low energy.
We find a unique naturally light vector- or scalar-like distortion field, which can be dynamical and have phenomenological implications. In particular, a light scalar particle that mixes with the Higgs boson
naturally appears.

\end{abstract}

\maketitle

%%%%%%%%%%%%%%%%%%%%%%%
\section{Introduction}
%%%%%%%%%%%%%%%%%%%%%%%

Einstein's theory of gravity, based on general covariance and the equivalence principle, has been highly successful as a correct description of gravitation. However, this is not the only possible formulation of gravity. In fact, in the Palatini formalism, where the metric and the connection are treated as independent variables, one often obtains a theory that is almost equivalent to Einstein gravity in the low energy limit. If a theory of gravity consistent with observations could be derived from fewer assumptions, that would clearly be preferable. In such a case, the theory might contain more dynamical degrees of freedom than Einstein gravity, and it should in principle be distinguishable from it. If phenomena beyond the Standard Model, such as inflation, dark matter, dark energy, or baryogenesis, are related to degrees of freedom residing in the gravitational sector, nothing could be more intriguing.

To this end, we consider the most general theory of gravity, i.e., \textit{metric-Affine gravity}~\cite{Hehl:1994ue}. We briefly summarize its essence below (see Appendix~\ref{sec:tor} for definitions of several quantities and e.g. Refs.~\cite{Iosifidis:2018jwu,Baldazzi:2021kaf} for reviews).
We introduce two quantities which are constructed from the Affine connection ${\Gamma^\rho}_{\mu\nu}$. One is the torsion tensor,
\begin{align}
	{T^\rho}_{\mu\nu} \equiv {\Gamma^\rho}_{\mu\nu} - {\Gamma^\rho}_{\nu\mu},
    \label{torsion}
\end{align}
and the other is the non-metricity tensor,
\begin{align}
	Q_{\rho\mu\nu} \equiv -\nabla_{\rho} g_{\mu\nu}.
    \label{nonmetricity}
\end{align}
Both torsion and non-metricity transform as correct tensors under diffeomorphism.
From these definitions, the Affine connection is expressed as
\begin{align}
	{\Gamma^\rho}_{\mu\nu} = \overset{\circ}{\Gamma}\,^\rho_{~\mu\nu} + {K^\rho}_{\mu\nu}, \label{Affine}
\end{align}
where $\overset{\circ}{\Gamma}\,^\rho_{~\mu\nu}$ is the Levi-Civita connection expressed in terms of the metric (\ref{LeviCivita}) and ${K^\rho}_{\mu\nu}$ is the so-called \textit{distortion} tensor, given by
\begin{align}
    {K^\rho}_{\mu\nu} = \frac{1}{2}\left( {T^\rho}_{\mu\nu} + T_{\nu~\mu}^{~\rho} -T_{\mu\nu}^{~~\rho}  +Q_{\mu\nu}^{~~\rho} + Q_{\nu~\mu}^{~\rho}  -Q_{~\mu\nu}^{\rho}\right).
    \nonumber
\end{align}
In this paper, we call $T^\rho_{~\mu\nu}$ and $Q^{\rho}_{~\mu\nu}$ ``distortion'' collectively.
In the conventional metric formulation of general relativity, both torsion and non-metricity are taken to be zero.
%However, it would be more attractive if we can start from fewer assumptions and arrive at a consistent theory. 
%The metric-Affine gravity is the most general formulation in this respect, since 
The metric-Affine gravity is the most general framework in a sense that both torsion and non-metricity are treated as independent variables.

The expression for the Riemann tensor is given in Eq.~(\ref{R_gK}) and the Ricci scalar is written as
\begin{align}
	R= \overset{\circ}{R}
	+\overset{\circ}\nabla_\mu K^{\mu\nu}_{~~\nu} - \overset{\circ}\nabla_\nu K^{~\mu\nu}_{\mu}
	+K^\mu_{~\mu\lambda}K^{\lambda\nu}_{~~\nu} - K^{\mu\nu}_{~~\lambda}K^\lambda_{~\mu\nu},
    \label{Ricci_K}
\end{align}
where $\overset{\circ}{R}$ denotes the Ricci scalar expressed in terms of the metric and $\overset{\circ}{\nabla}_\mu$ the covariant derivative with respect to the Levi-Civita connection, just as in the standard general relativity.
In a minimal setup, we only introduce the Einstein-Hilbert (EH) action as a gravity sector. Since $R$ is a function of both $g_{\mu\nu}$ and $K^\rho_{~\mu\nu}$, we need to take variations with respect to them independently.
As shown below, in this minimal setup none of the distortions is dynamical, and the resulting theory is almost the same as general relativity. 
However, there are lots of possibilities to add non-minimal terms and (some of) distortions can easily be dynamical.
It would be interesting if these distortion fields, originating from the gravitational sector, may take roles of explaining phenomena beyond the Standard Model.
Attempts have been made in this framework for inflation~\cite{BeltranJimenez:2015pnp,BeltranJimenez:2016wxw,Shimada:2018lnm,Aoki:2020zqm,Langvik:2020nrs,Mikura:2020qhc,Pradisi:2022nmh,Salvio:2022suk,Gialamas:2022xtt,Racioppi:2024zva} (see also Refs.~\cite{Shaposhnikov:2020gts,DiMarco:2023ncs,He:2024wqv,Gialamas:2024iyu,He:2025bli,Gialamas:2026pjo} in the framework of Einstein-Cartan gravity, in which the non-metricity is taken to be zero), dark matter physics~\cite{Barman:2019mlj,Shaposhnikov:2020aen,Rigouzzo:2023sbb}, and axion-like particle~\cite{Karananas:2025ews,He:2025fij} (though early attempts to address the strong CP problem~\cite{Mielke:2006zp,Mercuri:2009zi,Lattanzi:2009mg,Castillo-Felisola:2015ema,Karananas:2018nrj,Karananas:2024xja} do not work~\cite{Karananas:2025ews,He:2025fij}).
For such phenomenological purposes, a dynamical distortion is often introduced whose mass should be much smaller than the Planck scale. However, there are so many fields and possible terms in the action and hence it is necessary to just pick up small set of terms and make ad hoc assumptions on the parameters.
We will see that there is a unique vector distortion component that is naturally light, providing a well-motivated starting point for considering phenomenology from the gravitational sector.

In Sec.~\ref{sec:dis} we will see that there is a unique vector distortion mode that does not show up in the EH action. More generally, in a theory with projective invariance, this vector mode completely disappears from the action.
In Sec.~\ref{sec:dyn} it is shown that such a vector mode can be dynamical by imposing an approximate projective invariance and introducing a small projective-invariance-breaking term.
Depending on the model, a light dynamical scalar or vector fields appears. 
The smallness of its mass is ensured by the projective symmetry.
In Sec.~\ref{sec:sum} we summarize our results.

%%%%%%%%%%%%%%%%%%%%%%%%%%%%%%%%
\section{Distortion and projective invariance}
\label{sec:dis}
%%%%%%%%%%%%%%%%%%%%%%%%%%%%%%%%

In order to classify the degrees of freedom in the distortion sector, it is useful to decompose them into several vector and tensor parts. See Eqs.~(\ref{Tmu})-(\ref{Qmunu}) for the decomposition.
By substituting it to Eq.~(\ref{Ricci_K}), the Ricci scalar becomes
\begin{align}
	&R = \overset{\circ}{R} + \overset{\circ}{\nabla}_\mu(2T^\mu - Q^\mu + \hat Q^\mu) 
    +\frac{1}{24}\hat T_\mu \hat T^\mu \nonumber \\
    &-\frac{2}{3}T_\mu(T^\mu-Q^\mu + \hat Q^\mu) 
    -\frac{11}{72}Q_\mu Q^\mu + \frac{2}{9}Q_\mu \hat Q^\mu + \frac{\hat Q_\mu \hat Q^\mu}{18} \nonumber \\
    &+\frac{1}{2}\widetilde T_{\mu\nu\rho}\widetilde T^{\mu\nu\rho} - \widetilde T_{\mu\nu\rho} \widetilde Q^{\nu\rho\mu} + \frac{\widetilde Q_{\mu\nu\rho}}{4}(\widetilde Q^{\mu\nu\rho}-2 \widetilde Q^{\rho\mu\nu}).
    \label{Ricci_TQ}
\end{align}
This coincides with Ref.~\cite{Rigouzzo:2023sbb} (note that our sign convention of the non-metricity tensor is opposite).
We see that there are no mixings among pure tensor part $\widetilde T_{\mu\nu\rho}, \widetilde Q_{\mu\nu\rho}$ and vector parts.
Equations of motion for the tensor parts give constraint equations that are easily solved as $\widetilde T_{\mu\nu\rho}=0$ and $\widetilde Q_{\mu\nu\rho}=0$ as far as only the EH action is considered, so below we neglect them.
%Therefore, we just neglect the pure tensor part and focus on vector parts.
The terms with the Levi-Civita covariant derivative $\overset{\circ}{\nabla}_\mu$ do not contribute to the action after partial integration. 
The remaining terms are quadratic in vector quantities. 
As a result, the EH action is written in a very simple form:
\begin{align}
	S_{\rm EH}&=\int d^4x\sqrt{-g}\,\frac{M_{\rm Pl}^2}{2}R \nonumber \\
    &=\int d^4x\sqrt{-g}\,\left[\frac{M_{\rm Pl}^2}{2}\overset{\circ}{R} - \frac{1}{2}g^{\mu\nu}\,V_\mu \mathcal M^2 V_{\nu}^{\rm T}\right],
    \label{SEH_VMV}
\end{align}
where $V_\mu \equiv (\hat T_\mu,T_\mu, Q_\mu, \hat Q_\mu)$, $M_{\rm Pl}$ is the reduced Planck scale and
%%
%\begin{align}
%	\mathcal M^2 = M_{\rm Pl}^2\begin{pmatrix}
%      \displaystyle -\frac{1}{24} & \displaystyle 0 &\displaystyle 0   &\displaystyle 0 \\
%      \displaystyle 0 & \displaystyle \frac{2}{3} &\displaystyle -\frac{1}{3}   &\displaystyle \frac{1}{3} \\
%      \displaystyle 0 & \displaystyle-\frac{1}{3} &\displaystyle  \frac{11}{72}&\displaystyle -\frac{1}{9} \\
%      \displaystyle 0 & \displaystyle\frac{1}{3} &\displaystyle-\frac{1}{9}   &\displaystyle -\frac{1}{18}
%\end{pmatrix}.
%\label{massmatrix}
%\end{align}
%%
%%
\begin{align}
	\mathcal M^2 = M_{\rm Pl}^2\begin{pmatrix}
       \displaystyle -1/24 & \displaystyle 0 &\displaystyle 0   &\displaystyle 0 \\
       \displaystyle 0 & \displaystyle 2/3 &\displaystyle -1/3   &\displaystyle 1/3 \\
       \displaystyle 0 & \displaystyle-1/3 &\displaystyle 11/72&\displaystyle -1/9 \\
       \displaystyle 0 & \displaystyle 1/3 &\displaystyle -1/9   &\displaystyle -1/18
    \end{pmatrix}.
    \label{massmatrix}
\end{align}
The matrix $\mathcal M^2$ has eigenvalues $M_i^2$ $(i=1,2,3,4)$ of
\begin{align}
	M_i^2 = M_{\rm Pl}^2\left(-\frac{1}{24},~0,~\frac{55+\sqrt{6721}}{144},~\frac{55-\sqrt{6721}}{144}\right).
    \label{mass_eigen}
\end{align}
The first eigenvalue $M_1^2$ is trivial since $\hat T_\mu$ is decoupled from other vectors.
Remarkably, there is a zero eigenvalue.
The corresponding eigenstate, which we denote by $N_\mu$, is given by
\begin{align}
	N_\mu = \frac{1}{\sqrt{77}}\left(3T_\mu +8 Q_\mu + 2\hat Q_\mu \right).
    \label{Nmu}
\end{align}
The other two (unnormalized) eigenstates are
\begin{align}
	N_\mu^{\pm} = \frac{59\pm\sqrt{6721}}{54}T_\mu - \frac{95\pm\sqrt{6721}}{144}Q_\mu + \hat Q_\mu,
\end{align}
where plus and minus correspond to the eigenvalue $M_3^2$ and $M_4^2$, respectively.

We call $N_\mu$ and $N_\mu^{\pm}$ the ``massless'' and ``massive'' states, respectively, since the term quadratic in $V_\mu$ in the action (\ref{SEH_VMV}) looks like mass term.
However, it should be noticed that one should not regard it as a ``mass'' term without great care.
At the level of EH action, $V_\mu$ has no kinetic term and they are non-dynamical, and hence the quadratic term should not be interpreted as a mass term.
Interpretation becomes even more nontrivial if additional terms consisting of $V_\mu$ are included and dynamical degrees of freedom appear.
Sometimes the quadratic term becomes indeed a mass term, but sometimes the quadratic term may take a role of kinetic term rather than the mass term (see e.g. Ref.~\cite{He:2024wqv} for how the torsion $\hat T_\mu$ becomes dynamical in this context\footnote{
    For the kinetic term interpretation to be consistent, the corresponding eigenvalue in (\ref{mass_eigen}) must be negative. Otherwise it would become ghost. For the mass term interpretation to be consistent, the corresponding eigenvalue in (\ref{mass_eigen}) must be positive. Otherwise it would become tachyon.
}).
Nevertheless, we call $N_\mu$ a massless mode since we will actually see that $N_\mu$ can naturally be a dynamical field with small mass.

The existence of massless mode can be understood as a consequence of the projective symmetry. The projective transformation~\cite{Sandberg:1975db,Hehl:1978zkk} is given by
\begin{align}
	{\Gamma^\rho}_{\mu\nu} \to {\Gamma^\rho}_{\mu\nu} + \delta^\rho_{~\nu} \xi_\mu,
    \label{pro}
\end{align}
with an arbitrary space-time-dependent vector $\xi_\mu$. The Ricci scalar $R$ is invariant under the projective transformation, which is most easily seen in the general expression (\ref{R_Affine}).
The non-metricity and torsion are transformed as
\begin{align}
    &\hat Q^\mu\to \hat Q^\mu+2\xi^\mu,~~Q^\mu\to Q^\mu + 8\xi^\mu,~~\widetilde Q_{\rho\mu\nu}\to \widetilde Q_{\rho\mu\nu},\nonumber \\
	&\hat T^\mu \to \hat T^\mu,~~T^\mu \to T^\mu + 3\xi^\mu,~~\widetilde T_{\rho\mu\nu}\to \widetilde T_{\rho\mu\nu}.
    \label{pro_trans}
\end{align}
One can make two independent linear combinations from three vectors $T_\mu, Q_\mu, \hat Q_\mu$ so that they are invariant under the projective transformation, and define a unique vector that is orthogonal to both of these linear combinations. 
It is exactly the massless mode that we find, $N_\mu$.
Actually, one recognizes that the coefficients in (\ref{Nmu}) are aligned with the transformation law (\ref{pro_trans}) and it immediately means that the other two orthogonal eigenvectors $N_\mu^{\pm}$ are invariant under this transformation.
%In other words, $N_\mu$ may be regarded as a gauge mode with respect to the projective transformation.
Rewritten in terms of the eigenstate basis, $N_\mu$ does not appear at all in the action (\ref{Ricci_TQ}).
This is a starting point for considering non-trivial extended theories.
\\

%To summarize the discussion so far, a unique \textit{projective non-invariant mode} remains massless while the other orthogonal projective invariant modes become massive, as far as projective invariant theories (e.g. $f(R)$ theories~\cite{Sotiriou:2008rp,Olmo:2011uz}) are concerned.

%%
\begin{table*}[t]
\begin{center}
\begin{tabular}{|c|c|c|c|c|} \hline
    ~ & exact (EH) & exact (beyond EH) & approximate & none  \\ \hline\hline
    scalar projective symmetry & unphysical $N_\mu$ & massless vector $N_\mu$  & light vector $N_\mu$  & heavy vector $N_\mu$ \\ \hline
    %phenomenology: GR+$\dots$  & 4-Fermi & 4-Fermi + dark photon & 4-Fermi + dark photon & 4-Fermi\\ \hline \hline
    vector projective symmetry & unphysical $N_\mu$ & non-dynamical $N_\mu$ & light scalar $\nabla_\mu N^\mu$   & heavy scalar $\nabla_\mu N^\mu$ \\ \hline
    %phenomenology: GR+$\dots$  & 4-Fermi & 4-Fermi & 4-Fermi + inflaton/dark matter   & 4-Fermi \\ \hline
    %~         &        ~     & $N$ parity needed   & $N$ parity needed & ~ \\ \hline
\end{tabular}
\caption{The status of $N_\mu$ in various cases. ``Exact (EH)'' means that only the EH action is introduced, while ``exact (beyond EH)'' means that arbitrary terms consisting of distortion fields (as well as curvature terms) are introduced as far as they are consistent with the corresponding projective symmetry. For the scalar projective symmetry, the kinetic term for $N_\mu$ can be introduced so that it becomes dynamical. In the ``approximate'' column, we introduce (small) symmetry breaking terms. In the case of vector projective symmetry, we can introduce a term that makes $N_\mu$ dynamical at this stage. Only in the ``approximate'' case, nontrivial phenomenology appears. In the most other cases, the resulting theory just looks like general relativity plus Standard Model up to Planck-suppressed 4-Fermi interactions after integrating out distortions.}
\end{center}
\end{table*}
%%

%%%%%%%%%%%%%%%%%%%%%%%%%%%%%%%%%%%%%
\section{Naturally Light Distortion}
\label{sec:dyn}
%%%%%%%%%%%%%%%%%%%%%%%%%%%%%%%%%%%%%

%\paragraph{{\bf Making distortion dynamical.}}
We found a ``massless'' vector mode $N_\mu$ (\ref{Nmu}) in a projective invariant theory (it includes e.g. $f(R)$ theories~\cite{Sotiriou:2008rp,Olmo:2011uz}). Not only the mass term, but also kinetic term does not exist in the EH action and hence it is not even an auxiliary field: it is rather like a gauge mode that can be gauged away. 
On the other hand, all the other distortion modes are non-dynamical fields at this stage.
The equations of motion for distortions just give constraints for them, resulting in contact interactions among matter fields suppressed by the Planck scale, though details depend on how the matter sectors are coupled to distortion fields. As a result, the general relativity is effectively recovered up to Planck-suppressed contact terms.

More interesting situations happen when we introduce kinetic and mass terms for distortion vectors so that they become dynamical. However, the exact projective invariance under the transformation (\ref{pro}) forbids the kinetic term for $N_\mu$.
Here note that the projective symmetry is \textit{not} a requirement of a theory, but is just an emergent one if we restrict ourselves to the EH action. There is a priori no reason to demand the exact projective symmetry to the whole theory, but we want to still keep a partial benefit of it.
Thus we assume that there remains (approximate) restricted projective symmetry in which the transformation parameter $\xi_\mu$ is constrained under a certain condition.
We consider two options. (i) $\xi_\mu$ is written in the form of $\xi_\mu = \partial_\mu \xi$ with some scalar function $\xi$, which we call \textit{scalar projective symmetry}. (ii) $\xi_\mu$ is assumed to satisfy the condition $\overset{\circ}{\nabla}_\mu\xi^\mu=0$, which we call \textit{vector projective symmetry}.\footnote{
    In this paper we regard a resulting theory as a classical effective theory and we do not discuss full quantum treatments of theories with scalar/vector projective symmetry.
}
We again note that it is allowed to introduce any term consisting of projective invariant fields $\hat T_\mu, N^+_\mu, N^-_\mu, \widetilde T_{\mu\nu\rho}, \widetilde Q_{\mu\nu\rho}$, so they can become heavy fields though we need to take care of the appearance of ghosts~\cite{BeltranJimenez:2019acz,Aoki:2019rvi,BeltranJimenez:2020sqf,Iosifidis:2021xdx}. 
On the other hand, Ref.~\cite{Barker:2024dhb} proposed so-called extended projective symmetry, which may constrain allowed terms in the action.
Our model can be embedded into the framework of extended projective symmetry, since $N_\mu$ is not invariant anyway.

%%%%%%%%%%%%%%%%%%%%%%%%%%%%%%%%%%%%%
\subsection{Light vector distortion}
\label{sec:vec}
%%%%%%%%%%%%%%%%%%%%%%%%%%%%%%%%%%%%%

Let us first consider the case of approximate scalar projective symmetry, where $\xi_\mu$ is written in the form of $\xi_\mu = \partial_\mu \xi$.
As a minimal extension to the EH action, let us add an action for $N_\mu$ as $S_N=\int d^4x\sqrt{-g}\,\mathcal L_N$ where
%$S_N = S_{N,{\rm vector}}+ S_{N,{\rm mass}}$ where
%%
\begin{align}
    &\mathcal L_N = -\frac{1}{4}g^{\mu\nu}g^{\rho\sigma}N_{\mu\rho}N_{\nu\sigma}-\frac{1}{2}g^{\mu\nu} m^2N_{\mu}N_{\nu},
    \label{L_vector}
\end{align}
where $N_{\mu\nu} \equiv \overset{\circ}{\nabla}_\mu N_\nu-\overset{\circ}{\nabla}_\nu N_\mu$.
The first term is consistent with scalar projective symmetry, while the mass term violates it, ensuring the smallness of mass parameter $m$.
This is exactly the action of massive vector boson for $m^2>0$, providing a theoretical background for light dark photon studies.\footnote{
    A possibility of dark photon in metric-Affine gravity has been mentioned in Ref.~\cite{Pradisi:2022nmh} without specifying concrete realization of light state. See also Refs.~\cite{BeltranJimenez:2014iie,BeltranJimenez:2015pnp} for more general possibility of vector distortion. }
We emphasize that, in the most general metric-Affine gravity without projective invariance, there is a priori no reason to assume $m$ is hierarchically smaller than the Planck scale. Our framework gives a natural realization of light degrees of freedom in the distortion sector.

%A remark is that, the ``mass term'' does not always literally mean the mass term.
%For example, one can add a term like $\left(\nabla_\mu \hat T^\mu\right)^2$ to the minimal EH action. As shown in Ref.~\cite{He:2024wqv}, it reduces to an effective dynamical scalar field model and the term $\frac{1}{24}\hat T_\mu \hat T^\mu$ rather takes a role of kinetic term for this scalar.
%Still the natural mass scale is the Planck scale without a large/small parameter, as in the famous Starobinsky inflation model~\cite{Starobinsky:1980te}.

Let us discuss phenomenological implications of dynamical vector field $N_\mu$.
The only coupling of $N_\mu$ to the Standard Model fields at the renormalizable level consistent with the symmetry is the kinetic mixing with the Standard Model hypercharge photon,\footnote{
    In this paper we assume that all the matter fields are invariant under the projective transformation.
}
\begin{align}
    S_{N,{\rm mix}} = \int d^4x \sqrt{-g}\left[ \frac{\epsilon}{2}g^{\mu\nu}g^{\rho\sigma}N_{\mu\rho}B_{\nu\sigma}\right].
    \label{kinmix}
\end{align}
where $B_{\nu\sigma}$ is the field strength tensor of the hypercharge photon and $\epsilon$ is the mixing parameter. After the electroweak symmetry breaking, we obtain the kinetic mixing between $N_\mu$ and the Standard Model massless photon.
However, there are strong constraints on the parameter $\epsilon$, as summarized in Ref.~\cite{Caputo:2021eaa} for $m\lesssim 100\,$keV and in Ref.~\cite{Fabbrichesi:2020wbt} for $m\lesssim 1\,$TeV.
%\footnote{
%In the exactly massless limit $m=0$, the constraint disappears since the kinetic mixing is completely removed by the field redefinition without leaving any change in the remaining action.}
Therefore, in the broad mass range $10^{-22}\,{\rm eV} \lesssim m \lesssim 1$\,TeV, an extra mechanism is needed to suppress the size of $\epsilon$.
A simple prescription is to assign a parity (or charge conjugation) symmetry, under which only $N_\mu$ changes its sign while all other fields do not. Then $\epsilon$ may be regarded as a small parity breaking parameter.
The light dark photon with $m\gtrsim 1$\,$\mu$eV with a small kinetic mixing is a good dark matter candidate, since inflationary fluctuations or gravitational production works as a correct production mechanism~\cite{Graham:2015rva,Ema:2019yrd,Ahmed:2020fhc}.

%%%%%%%%%%%%%%%%%%%%%%%%%%%%%%%%%%%%%
\subsection{Light scalar distortion}
\label{sec:scalar}
%%%%%%%%%%%%%%%%%%%%%%%%%%%%%%%%%%%%%

Next let us consider the case of approximate vector projective symmetry, where $\xi^\mu$ satisfies $\overset{\circ}{\nabla}_\mu\xi^\mu=0$.
In this case, as a minimal extension to the EH action, we can write the action for $N_\mu$ as
\begin{align}
    \mathcal L_N=\frac{1}{2}\left(\overset{\circ}{\nabla}_\mu N^\mu\right)^2 +\frac{1}{2}g^{\mu\nu} m^2N_{\mu}N_{\nu},
    \label{L_scalar}
\end{align}
The first term is consistent with the symmetry, while the second term violates it ensuring the smallness of $m$, similar to the previous case.
This action can be rewritten by introducing an auxiliary scalar field $\chi$ as
\begin{align}
    S_N &= \int d^4x \sqrt{-g}\left[\frac{m^2}{2}N_\mu N^\mu +\frac{1}{2}\left( 2\chi\overset{\circ}{\nabla}_\mu N^\mu- \chi^2\right)\right] \nonumber\\
    &= \int d^4x \sqrt{-g}\left[ -\frac{1}{2}g^{\mu\nu}\partial_\mu\varphi \partial_\nu\varphi  -\frac{m^2}{2}\varphi^2\right],
\end{align}
where $\varphi\equiv \chi/m$ is the canonically normalized scalar. In the second line we have solved the equation of motion for $N_\mu$ and substituted it to the action.
Note that the sign of the mass term in (\ref{L_scalar}) is opposite to (\ref{L_vector}) so that the kinetic term of the scalar has a correct sign.
Thus it reduces to a theory of just a massive scalar.

The only coupling to the Standard Model matter consistent with the symmetry at the renormalizable level is\footnote{
    \label{foot:current}
    Terms like $\mathcal L \sim N^\mu J_\mu$ where $J_\mu$ denotes the gauge invariant current consisting of fermions or Higgs are not invariant under the vector projective symmetry. 
}
\begin{align}
    S_{N,{\rm mat}} = \int d^4x \sqrt{-g}\,N^\mu\left(c_H\partial_\mu |H|^2\right),
\end{align}
where $H$ denotes the Higgs doublet and $c_H$ is the coupling constant.
Again by using the equation of motion of $N_\mu$ and substituting it to the action, we find the total action as
\begin{align}
   \mathcal L_N = -\frac{1}{2}g^{\mu\nu}\partial_\mu\varphi \partial_\nu\varphi
    -\frac{1}{2}\left(m\varphi + c_H|H|^2\right)^2,
    \label{cH}
\end{align}
where $\varphi \equiv (\chi-c_H|H|^2)/m$. 

Let us mention some phenomenological implications of this massive scalar.
If $m\sim 10^{13}$\,GeV, it may be identified as the inflaton (after modifying the potential to fit to the observational data). See Ref.~\cite{He:2024wqv} for a related idea in the Einstein-Cartan framework.
In our construction, the smallness of inflaton mass, $m \ll M_{\rm Pl}$, is naturally explained.
The form of the matter interaction indicates that the inflaton can decay into the Higgs particles, which leads to successful reheating.

\begin{figure}[t]
  \centering
  \includegraphics[width=0.45\textwidth]{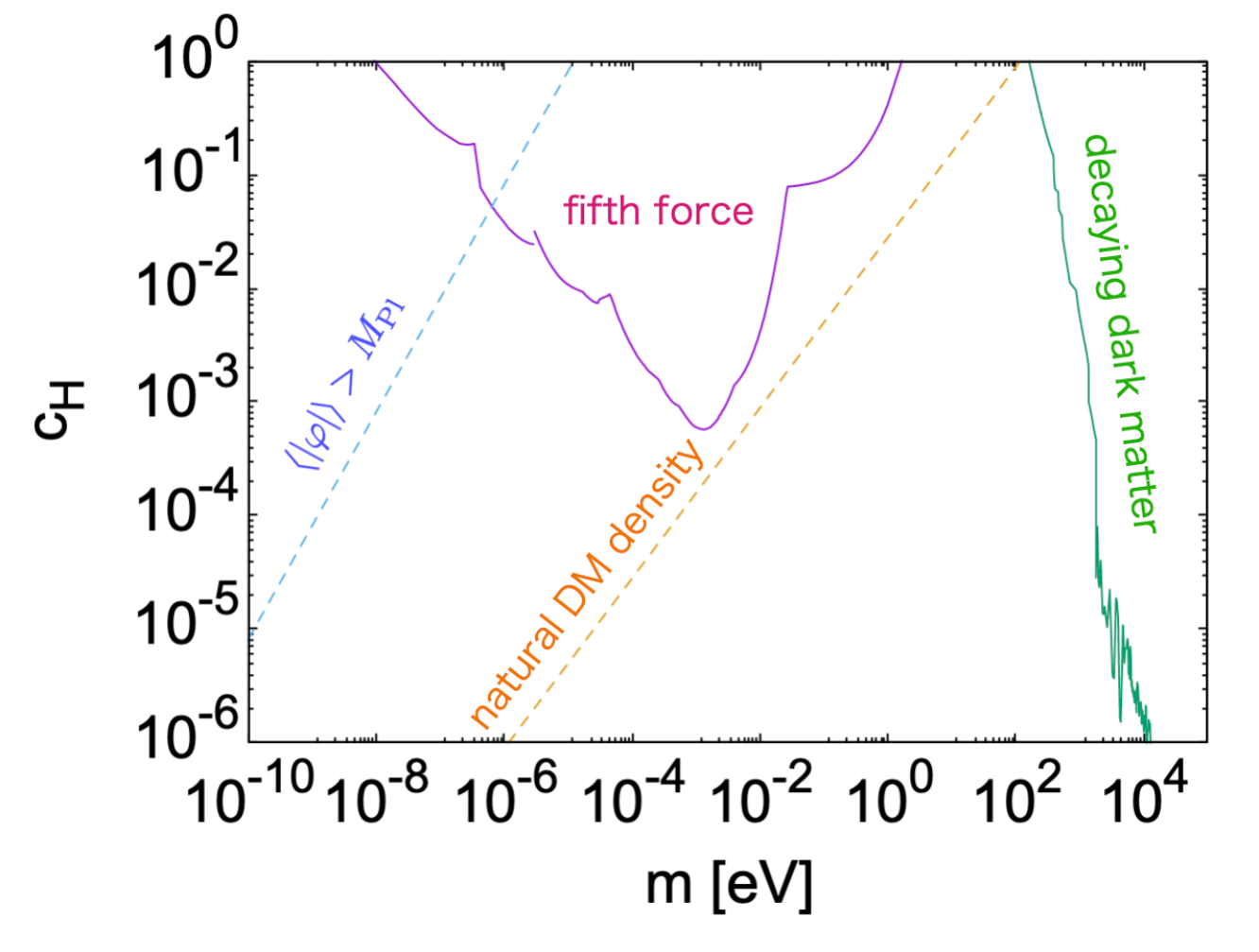}
  \caption{Constraints on $c_H$ in (\ref{cH}) as a function of mass of $\varphi$. The region above the purple line is excluded by the fifth force experiment and the region above the green line is mostly excluded by X-ray observations assuming $\varphi$ is total dark matter. 
  On the orange dashed line labeled as ``natural DM density'', the dark matter abundance is naturally explained by the misalignment mechanism, though it is still possible to obtain correct dark matter abundance in other regions depending on the initial field value.  
  We also show the region with $\langle|\varphi|\rangle > M_{\rm Pl}$ for reference.}
  \label{fig:cH}
\end{figure}

As another possible implication, $\varphi$ may be identified as dark matter.
A characteristic property of $\varphi$ is that it mixes with the Higgs. The mixing angle is $\theta \simeq c_H mv/m_h^2$, where $v=246\,$GeV is the Higgs vacuum expectation value (VEV) and $m_h=125$\,GeV is the Higgs mass.
%First note that $\varphi$ develops a VEV of $\langle|\varphi|\rangle\sim c_Hv^2/m$ and if we demand that it should not exceed $M_{\rm Pl}$, we obtain a lower bound on $m$ as $m \gtrsim 10^{-5}\,$eV $\times c_H$.
Due to the mixing, $\varphi$ can decay into two photons (see Appendix~\ref{sec:decay} for the calculation of the decay rate). In order for its lifetime to be longer than the age of the Universe, $m \lesssim 10^4$\,eV $\times c_H^{-2/5}$ is required, although observations of Galactic/extragalactic photons give more stringent constraints~\cite{Cadamuro:2011fd}. 
Another constraint comes from the fifth force, since $\varphi$ mediates a long range force through the mixing with Higgs. The effective coupling with the fermion is written as\footnote{
    One may also introduce terms like $N_\mu J^\mu\propto(\partial_\mu\varphi)J^\mu$ (see footnote~\ref{foot:current}).
    But this term is not very relevant for several reasons. First, the coefficient of this term is at most $\sim1/M_{\rm Pl}$ considering the violation of vector projective symmetry. Second, its contribution to long range force is negligible, since upon integration by parts it is proportional to $\partial_\mu J^\mu$ and it vanishes for on-shell fermions if $J^\mu$ is vector current. If $J^\mu$ is axial, it gives spin-dependent force but the constraint is much weaker. 
}
\begin{align}
     \mathcal L \simeq c_H \frac{m \varphi}{m_h^2} \sum_\psi m_\psi \overline\psi\psi = \frac{c_H^2 v^2}{2m_h^2}\frac{\varphi}{\langle|\varphi|\rangle}  \sum_\psi m_\psi \overline\psi\psi,
     \label{L_phipsi}
\end{align}
where $\langle|\varphi|\rangle \sim c_H v^2/(2m)$ is the VEV of $\varphi$.
For very wide mass range of $10^{-22}\,{\rm eV} \lesssim m \lesssim 10^{-3}\,{\rm eV}$, the new force must be (much) smaller than the gravitational one~\cite{Adelberger:2003zx,KONOPLIV2011401}. 
Fig.~\ref{fig:cH} shows constraints on the parameters $(m,c_H)$. The region above the purple line is excluded by the fifth force experiments~\cite{Adelberger:2003zx,Schlamminger:2007ht} and the region above the green line is mostly excluded by X-ray observations assuming $\varphi$ is total dark matter~\cite{Cadamuro:2011fd}.
For drawing these lines, we used the data available in Ref.~\cite{AxionLimits}.
We also show the region with $\langle|\varphi|\rangle > M_{\rm Pl}$ just for reference, but this may not be regarded as a constraint since the VEV of $\varphi$ can be absorbed by the tadpole renormalization.
Note that our model is radiatively stable since the coupling is suppressed by the scalar mass $m$ itself. 
%We see that huge parameter regions are available for dark matter $\varphi$.

%%%%%%%%%%%%%%%%%%%%%%%%%%%%%%%%%%%%%%%%%%%%%%%%%%%%%%%%%%%%%%%%%%%%
%%%%%%%%%%%%%%%%%%%%%%%%%%%%%%%%%%%%%%%%%%%%%%%%%%%%%%%%%%%%%%%%%%%%%

Let us estimate the scalar abundance through conventional misalignment mechanism. The energy-to-entropy-density ratio is given by
\begin{align}
	\frac{\rho_\phi}{s}\simeq \frac{T_{\rm osc}}{8} \left(\frac{\varphi_{\rm ini}}{M_{\rm Pl}}\right)^2,
\end{align}
where $T_{\rm osc}$ denotes the temperature at which $\varphi$ begins to oscillate, which is roughly given by $T_{\rm osc}\sim \sqrt{m M_{\rm Pl}}$ (assuming that the reheating temperature is so high that the oscillation starts at the radiation-dominated era). 
The initial amplitude $\varphi_{\rm ini}$ is a free parameter set by the initial condition, but a natural value may be $\varphi_{\rm ini} \sim \left<\varphi\right>=-c_H v^2/(2m)$ unless we fine-tune the initial condition. 
In this case we can estimate the abundance as
\begin{align}
	\frac{\rho_\phi}{s}\simeq 10^{-9}\,{\rm GeV}\left(\frac{c_H}{10^{-4}}\right)^2 \left(\frac{10^{-3}\,{\rm eV}}{m}\right)^{3/2}.
\end{align}
It can explain the observed dark matter abundance for some parameter regions, as shown in the Figure. 
Still, however, we should note that even in other parameter regions the dark matter abundance can be adjusted by choosing an initial displacement, as is often assumed in light scalar dark matter scenario.
Thus one should not interpret this as an excluded region.
Another constraint comes from isocurvature perturbation caused by the quantum fluctuation of $\varphi$ itself, and it gives upper bound on the inflationary Hubble scale $H_{\rm inf}$ as $H_{\rm inf}/\varphi_{\rm ini}\lesssim 10^{-5}$.

%%%%%%%%%%%%%%%%%%%%%%%%%%%%%%%%%%%%%
\section{Summary and discussion}
\label{sec:sum}
%%%%%%%%%%%%%%%%%%%%%%%%%%%%%%%%%%%%%

%\paragraph{{\bf Summary and discussion.}}
In the most general formulation of gravity, i.e., metric-Affine gravity, we found a unique vector distortion degree of freedom, whose lightness is technically natural in the 't Hooft sense~\cite{tHooft:1979rat}.
Depending on whether we postulate scalar or vector projective symmetry, there appears a light vector or scalar field, both of which have interesting phenomenological consequences.
In the case of vector-like distortion, it looks like dark photon, which may be kinetically mixed with the ordinary photon. Thus we provided a possibility that the origin of dark photon may be related to the spacetime geometry. 
In the case of scalar-like distortion, we found a characteristic property: it weakly couples to matter only through the mixing with Higgs and the mixing angle is suppressed by the scalar mass itself.
It may give a natural realization of the scenario considered in Ref.~\cite{Piazza:2010ye} and partly resembles the scenario of relaxion~\cite{Graham:2015cka}, in which a light scalar is mixed with the Higgs~\cite{Flacke:2016szy}.
At more philosophical level, it had been a bit strange that we eventually end up with the Einstein's general relativity even if we started from much weaker assumptions (i.e., allowing torsion and non-metricity), unless some ad hoc assumptions on the parameters are imposed. 
Our findings indicate that the deviation from the Einstein's general relativity can be rather natural due to the existence of light distortion mode.

%%%%%%%%%%%%%%%%%% Acknowledgements %%%%%%%%%%%%%%%%%%%%%%%%%%%%%%%%%%%%%%%
%\section*{Acknowledgments}
%%%%%%%%%%%%%%%%%%%%%%%%%%%%%%%%%%%%%%%%%%%%%%%%%%%%%%%%%%%%%%%%%%%%%%%%%%%

~\\
\textit{Acknowledgments.} This work was supported by JSPS KAKENHI (Grant Numbers 24K07010, 26K00695, 26H00403).
This work was also supported by World Premier International Research Center Initiative (WPI), MEXT, Japan.

%%% End Matter %%%
%\pagebreak
%\setcounter{equation}{0}
%\setcounter{figure}{0}
%\setcounter{table}{0}
%\makeatletter
%\renewcommand{\theequation}{E.\arabic{equation}}
%\renewcommand{\thefigure}{E.\arabic{figure}}
%\renewcommand{\bibnumfmt}[1]{[E#1]}
%\renewcommand{\citenumfont}[1]{E#1}
%\onecolumngrid
%\begin{center}
%\textbf{\large END MATTER}
%\end{center}
%\twocolumngrid
%%%

%%%%%%%%%%%%%%%%%%%%%%%%%%%%%%%%%%%%%%%%%%%%%%%%%%%%%%%%%%%
\appendix
%%%%%%%%%%%%%%%%%%%%%%%%%%%%%%%%%%%%%%%%%%%%%%%%%%%%%%%%%%%

%%%%%%%%%%%%%%%%%%%%%%%%%%%%%%%%%%%%%%%
\section{Connection and distortion}
\label{sec:tor}
%%%%%%%%%%%%%%%%%%%%%%%%%%%%%%%%%%%%%%%

We define the covariant derivative of some tensor $V$ by
\begin{align}
	&\nabla_\rho V^{\mu_1\mu_2\cdots}_{~~~\nu_1\nu_2\cdots} = \partial_\rho V^{\mu_1\mu_2\cdots}_{~~~\nu_1\nu_2\cdots} \nonumber\\
	&+\sum_i \Gamma^{\mu_i}_{~\rho\lambda} V^{\cdots \mu_{i-1} \lambda \mu_{i+1}\cdots}_{~~~~~\nu_1\nu_2\cdots} 
	-\sum_i \Gamma^{\lambda}_{~\rho\nu_i} V^{\mu_1\mu_2\cdots}_{~\cdots  \nu_{i-1} \lambda \nu_{i+1} \cdots},
    \nonumber
\end{align}
with the Affine connection $\Gamma^{\lambda}_{~\mu\nu}$, so that the covariant derivative has a correct tensor properties under diffeomorphism. The torsion tensor $T^\rho_{~\mu\nu}$ and the non-metricity tensor $Q_{\rho\mu\nu}$ are defined by Eqs.~(\ref{torsion}) and (\ref{nonmetricity}), respectively. The Riemann tensor is defined through
\begin{align}
	[\nabla_\mu, \nabla_\nu]V^\alpha \equiv R^\alpha_{~\beta\mu\nu}V^\beta - T^\beta_{~\mu\nu} \nabla_\beta V^\alpha,
    \label{Riemann_def}
\end{align}
for arbitrary vector $V^\alpha$. The Riemann tensor is constructed only from the Affine connection as
\begin{align}
    %&R^{\alpha}_{~\beta\mu\nu}(\Gamma) = \partial_\mu\Gamma^\alpha_{~\nu\beta} -  \partial_\nu\Gamma^\alpha_{~\mu\beta}+\Gamma^\alpha_{~\mu\lambda}\Gamma^\lambda_{~\nu\beta} - \Gamma^\alpha_{~\nu\lambda}\Gamma^\lambda_{~\mu\beta}. \nonumber\\
    R^{\alpha}_{~\beta\mu\nu}(\Gamma) = \left(\partial_\mu\Gamma^\alpha_{~\nu\beta} 
	+\Gamma^\alpha_{~\mu\lambda}\Gamma^\lambda_{~\nu\beta}\right) - (\mu\leftrightarrow\nu).
    \label{R_Affine}
\end{align}
One can substitute ${\Gamma^\rho}_{\mu\nu} = \overset{\circ}{\Gamma}\,^\rho_{~\mu\nu} + {K^\rho}_{\mu\nu}$ (Eq.~(\ref{Affine})), where the Levi-Civita connection is given by
\begin{align}
	\overset{\circ}{\Gamma}\,^\rho_{~\mu\nu} \equiv \frac{1}{2}g^{\rho\sigma}\left( \partial_\mu g_{\nu\sigma} + \partial_\nu g_{\sigma\mu}-\partial_{\sigma}g_{\mu\nu} \right),
    \label{LeviCivita}
\end{align}
to separate the Riemann tensor into the Levi-Civita part and the remaining part consisting of the distortion as
\begin{align}
	R^{\alpha}_{~\beta\mu\nu}(g, K) &= \overset{\circ}{R}\,^{\alpha}_{~\beta\mu\nu}(g)
	+\overset{\circ}\nabla_\mu K^\alpha_{~\nu\beta} - \overset{\circ}\nabla_\nu K^\alpha_{~\mu\beta} \nonumber\\
	&~~~~~~+K^\alpha_{~\mu\lambda}K^\lambda_{~\nu\beta} - K^\alpha_{~\nu\lambda}K^\lambda_{~\mu\beta}.
	\label{R_gK}
\end{align}
The Ricci scalar is given by $R = g^{\mu\nu}R^{\alpha}_{~\mu\alpha\nu}$.
In order to simplify the EH action, we further decompose the distortion (torsion and non-metricity). 

From the definition of distortion, we find
\begin{align}
	T^\rho_{~\mu\nu}=K^{\rho}_{~\mu\nu}-K^{\rho}_{~\nu\mu},~~~
    Q^\rho_{~\mu\nu}=K^{~\rho}_{\mu~\nu}+K^{~\rho}_{\nu~\mu}.
\end{align}
By noting that the torsion tensor $T_{\rho\mu\nu}$ is antisymmetric under $\mu\leftrightarrow\nu$ by definition, it is conveniently decomposed into two vectors and one pure tensor as 
\begin{align}
	&\hat T^\mu \equiv \epsilon^{\mu\nu\rho\sigma} T_{\nu\rho\sigma}, \label{Tmu}\\
	&T^\mu \equiv g_{\rho\sigma} T^{\rho\mu\sigma}, \\
	&\widetilde T_{\rho\mu\nu} \equiv T_{\rho\mu\nu}+\frac{1}{3}(g_{\rho\mu}T_\nu-g_{\rho\nu}T_{\mu})-\frac{1}{6}\epsilon_{\rho\mu\nu\sigma}\hat T^\sigma,
\end{align}
where the pure tensor part $\widetilde T_{\rho\mu\nu}$ satisfies $\epsilon^{\lambda\rho\mu\nu} \widetilde T_{\rho\mu\nu}=0$ and $g_{\rho\nu} \widetilde T^{\rho\mu\nu} = 0$. Note that it also satisfies
\begin{align}
	\widetilde T_{\rho\mu\nu} + \widetilde T_{\nu\mu\rho} + \widetilde T_{\mu\nu\rho} = 0
    ~\to~\widetilde T_{\rho\mu\nu} \widetilde T^{\rho\mu\nu} + 2\widetilde T_{\rho\mu\nu} \widetilde T^{\mu\nu\rho}=0. \nonumber
\end{align}
Similarly, the non-metricity tensor $Q_{\rho\mu\nu}$ is symmetric under $\mu\leftrightarrow\nu$ and it is decomposed as
\begin{align}
	& Q^\mu \equiv g_{\rho\sigma}Q^{\mu\rho\sigma},\\
	& \hat Q^\mu \equiv g_{\rho\sigma}Q^{\rho\mu\sigma},\\
	&\widetilde Q_{\rho\mu\nu}\equiv Q_{\rho\mu\nu}-\frac{1}{18}\left[g_{\mu\nu}(5Q_\rho-2\hat Q_\rho) \right.\nonumber\\
    &\left.~~~+ 4(g_{\rho\mu}\hat Q_\nu +g_{\rho\nu}\hat Q_\mu) - (g_{\rho\mu}Q_\nu +g_{\rho\nu}Q_\mu) \right].
    \label{Qmunu}
\end{align}
where the pure tensor part $\widetilde Q^{\rho\mu\nu}$ satisfies $g_{\mu\nu}\widetilde Q^{\rho\mu\nu}=0$ and $g_{\rho\nu}\widetilde Q^{\rho\mu\nu}=0$.
By substituting these decompositions into Eq.~(\ref{R_gK}), after straightforward but tedious calculations, we end up with the expression (\ref{Ricci_TQ}).

%%%%%%%%%%%%%%%%%%%%%%%%%%%%%%%%%%%%%%%
\section{Distortion-matter couplings}
\label{sec:coup}
%%%%%%%%%%%%%%%%%%%%%%%%%%%%%%%%%%%%%%%

Here we summarize how the distortion tensor can couple to matter fields minimally.
We mainly focus on the kinetic terms for a scalar, fermion and vector boson, in which the distortion may naturally appear as a part of covariant derivative, and see whether the projective invariance is naturally satisfied or not.
Recall that distortion vectors $\hat T_\mu, T_\mu,Q_\mu, \hat Q_\mu$ are classified into the projective non-invariant vector $N_\mu$ and the projective invariant ones $(N_\mu^\pm, \hat T_\mu)$. Below we sometimes use a notation $P_\mu$ to collectively represent any linear combination of projective invariant vectors $(N_\mu^\pm, \hat T_\mu)$.

Our convention of the $\gamma$ matrix is
\begin{align}
	\gamma^\mu = \begin{pmatrix} 0 & \sigma^\mu \\ \overline \sigma^\mu & 0 \end{pmatrix},
    ~~~\gamma_5 \equiv i\gamma^0\gamma^1\gamma^2\gamma^3 = \begin{pmatrix} 1 & 0 \\ 0 & -1 \end{pmatrix},
    \nonumber
\end{align}
where $\sigma^\mu \equiv (-1, \vec\sigma)$ and $\overline\sigma^\mu \equiv (-1,-\vec\sigma)$. The anti-commutation relation is $\{ \gamma^\mu, \gamma^\nu \} = -2\eta^{\mu\nu}$ with $\eta^{\mu\nu}={\rm diag}\,(-,+,+,+)$.
A useful formula for our purpose is
\begin{align}
	%&\gamma^{[\mu}\gamma^\nu\gamma^{\rho]} = -i\epsilon^{\mu\nu\rho\sigma} \gamma_\sigma \gamma_5, \\
    \gamma^{\mu}\gamma^\nu\gamma^{\rho} = g^{\rho\mu}\gamma^\nu - g^{\mu\nu}\gamma^\rho - g^{\nu\rho}\gamma^\mu
    - i \epsilon^{\mu\nu\rho\sigma}\gamma_\sigma\gamma_5. \label{3gamma}
\end{align}
The Levi-Civita tensor is defined with $\epsilon^{0123}=1$. 
%The symmetrization of the tensor index is defined as e.g. $\gamma^{[\mu}\gamma^{\nu]} =\frac{1}{2}(\gamma^\mu\gamma^\nu-\gamma^\nu\gamma^\mu)$.

\subsection{Scalar}

In the Standard Model, the Higgs doublet is only a fundamental scalar. Let $\Phi$ be a general complex scalar charged under some gauge symmetry. (We suppress the gauge index for notational simplicity.)  
In a flat space, the gauge invariant kinetic term is written as
\begin{align}
   \int d^4x\left[-(D_\mu \Phi)^\dagger(D^\mu \Phi)\right]
    =\int d^4x \left[\frac{1}{2}\Phi^\dagger D_\mu D^\mu \Phi + {\rm h.c.}\right],
    \label{Phi_kin}
\end{align}
where $D_\mu$ denotes the gauge covariant derivative in a flat space.
There are two natural prescriptions to extend it to the curved space: replace $D_\mu \to \nabla_\mu$ or $D_\mu \to \overset{\circ}{\nabla}_\mu$.\footnote{
    Note that both $\nabla_\mu$ and $\overset{\circ}{\nabla}_\mu$ has a correct transformation property as a vector under general coordinate transformation. Thus the requirement of diffeomorphism invariance does not restrict which derivative must be used.
}
The latter option leads to rather trivial consequences that there are no couplings between scalar and distortion, so we focus on the former option in the following.
The same caution applies to the case of fermion and vector boson discussed below.

But just promoting $D_\mu\to \nabla_\mu$ is also non-trivial, since the resulting theory is different depending on whether we start from the left hand side or right hand side of (\ref{Phi_kin}).
Clearly, the form of left hand side does not include the connection even in curved space while the right hand side does. 
One more complexity arises by noting that $\nabla_\mu \nabla^\mu$ and $\nabla^\mu \nabla_\mu$ do not commute in the presence of non-metricity:
\begin{align}
	(\nabla_\mu \nabla^\mu - \nabla^\mu \nabla_\mu) \Phi = \hat Q^\mu \nabla_\mu \Phi.
\end{align}
Thus the most general kinetic term is characterized by two real parameters $\alpha$ and $\beta$ as
\begin{align}
	S &= \int d^4x \sqrt{-g} \left\{-(1-\beta)g^{\mu\nu}(\nabla_\mu\Phi)^\dagger(\nabla_\nu\Phi) \right.\nonumber\\
    &\left. ~~+\frac{\beta}{2}\left[(1-\alpha)\Phi^\dagger\nabla_\mu\nabla^\mu\Phi + \alpha\Phi^\dagger\nabla^\mu\nabla_\mu\Phi + {\rm h.c.}\right] \right\} \nonumber\\
    &=\int d^4x \sqrt{-g} \left[-g^{\mu\nu}(\nabla_\mu\Phi)^\dagger(\nabla_\nu\Phi) \right.\nonumber\\
    & ~~~~~~~ \left. -\frac{\beta}{2}\left(T^\mu-\frac{Q_\mu}{2} + \alpha \hat Q_\mu \right)\partial_\mu|\Phi|^2 \right].
\end{align}
The projective symmetry is recovered for $\beta=0$ or $\alpha=1/2$~\cite{Shimada:2018lnm}. In this case there is no scalar-$N_\mu$ coupling. 
For more general values of $\alpha$ and $\beta$, couplings arise in the form of $\sim P^\mu \partial_\mu|\Phi|^2$ or $\sim N^\mu \partial_\mu|\Phi|^2$. As noted in Sec.~\ref{sec:scalar}, a term like $\sim N^\mu \partial_\mu|\Phi|^2$ is not invariant under the general projective transformation, but it is invariant under the vector projective transformation.

In addition, one can also introduce a term like $\sim P^\mu J_\mu$ or $\sim N^\mu J_\mu$ where $J_\mu = i(\Phi^\dagger\nabla_\mu\Phi-\Phi\nabla_\mu\Phi^\dagger)$, although the latter is not projective invariant.

\subsection{Fermion}

The covariant derivative of a fermion $\psi$ is defined through the spin connection $\omega_\mu^{AB}$ as
\begin{align}
	\nabla_\mu \psi = D_\mu \psi -\frac{1}{8}\omega_{\mu}^{AB}(\gamma_A\gamma_B-\gamma_B\gamma_A)\psi,
\end{align}
where $D_\mu$ is the gauge covariant derivative in the flat space.
The projective transformation acts as $\omega_{\mu~B}^{~A} \to \omega_{\mu~B}^{~A} +\delta^{A}_{~B}\xi_\mu$ and hence the covariant derivative has projective invariance.
By using a formula (use Eq.~(\ref{3gamma}) to derive this)
\begin{align}
	\gamma^\mu \nabla_\mu\psi = \gamma^\mu\left[ \overset{\circ}\nabla_\mu \psi+\frac{1}{8}\left(
    i\hat T_\mu \gamma_5 + 4T_\mu  - 2Q_\mu+ 2\hat Q_\mu\right)\psi\right],
    \nonumber
\end{align}
we obtain the fermion kinetic term as
\begin{align}
\frac{i}{2}\overline\psi \gamma^\mu \nabla_\mu \psi + {\rm h.c.}=  \left(\frac{i}{2}\overline\psi \gamma^\mu \overset{\circ}{\nabla}_\mu \psi +{\rm h.c.}\right)
	-\frac{1}{8}\hat T_\mu\,\overline\psi\gamma_5\gamma^\mu \psi,
    \nonumber
\end{align}
%%
%(Note that $i\overline\psi \gamma^\mu \nabla_\mu \psi$ itself is not Hermite in the curved space with distortion.)
In terms of the chiral fermion with left (L) or right (R) handed chirality, as in the case of Standard Model fermions, the kinetic term becomes
\begin{align}
\frac{i}{2}\overline\psi_{\bullet} \gamma^\mu \nabla_\mu \psi_{\bullet} + {\rm h.c.}=  \left(\nabla_\mu\to\overset{\circ}{\nabla}_\mu\right)
	\pm\frac{1}{8} \hat T_\mu\,\overline\psi_{\bullet}\gamma^\mu \psi_{\bullet}, \nonumber
\end{align}
where ${\bullet}=$ L or R and the plus (minus) sign applies to L (R).
Thus a fermion naturally couples to the torsion vector $\hat T_\mu$.
Instead one could have defined the fermion kinetic term through $\overset{\circ}{\nabla}_\mu$ from the first. In such a case, one obtains no fermion-torsion coupling.
Note also that the following term does not vanish in the presence of distortion~\cite{Freidel:2005sn,Rigouzzo:2023sbb}:
\begin{align}
\frac{1}{2}\overline\psi_{\bullet} \gamma^\mu \nabla_\mu \psi_{\bullet} + {\rm h.c.}
	=\left(T_\mu-\frac{Q_\mu}{2}+\frac{\hat Q_\mu}{2}\right)\overline\psi_{\bullet}\gamma^\mu \psi_{\bullet}.
    \nonumber
\end{align}
Thus we define the general fermion kinetic term as follows:
\begin{align}
    S &= \int d^4x \sqrt{-g} \left[
        \frac{i}{2}(1-i\alpha)\overline\psi_{\bullet} \gamma^\mu \nabla_\mu \psi_{\bullet} + {\rm h.c.}
    \right] \nonumber\\
    &=\int d^4x \sqrt{-g} \left\{ 
         \left(\frac{i}{2}\overline\psi_{\bullet} \gamma^\mu \nabla_\mu \psi_{\bullet} + {\rm h.c.}\right)\right. \nonumber \\
    &\left.+ \left[\pm\frac{1}{8}\hat T_\mu + \alpha \left(T_\mu-\frac{Q_\mu}{2}+\frac{\hat Q_\mu}{2}\right)\right]\overline\psi_{\bullet}\gamma^\mu\psi_{\bullet} \right\},
\end{align}
where $\alpha$ is a real parameter.
We soon notice that the both $\hat T_\mu$ and the combination $T_\mu-\frac{Q_\mu}{2} + \frac{\hat Q_\mu}{2}$ are projective invariant, and hence $N_\mu$ does not couple to fermions in this generalized kinetic term.

In addition, one can introduce terms like $\sim P^\mu J_\mu$ by hand without breaking projective invariance, where $J_\mu$ is gauge-invariant vector/axial current consisting of fermions.
If strict projective invariance is not required, one can also introduce a term like $\sim N^\mu J_\mu$.

\subsection{Vector}

Lastly let us consider a kinetic term for a gauge boson. First we consider a U(1) gauge boson $A_\mu$. In the presence of distortion, a naive substitution $\overset{\circ}{\nabla}_\mu\to\nabla_\mu$ in the field strength tensor leads to  
\begin{align}
	\nabla_\mu A_\nu - \nabla_\nu A_\mu = \partial_\mu A_\nu - \partial_\nu A_\mu -T^\rho_{~\mu\nu} A_\rho.
\end{align}
The last term breaks the gauge invariance, and hence we are forced to define the field strength tensor through $\overset{\circ}{\nabla}_\mu$. The resulting action is
\begin{align}
    S = \int d^4x \sqrt{-g}\left(-\frac{1}{4}g^{\mu\rho}g^{\nu\sigma}F_{\mu\rho}F_{\nu\sigma}\right),
\end{align}
where $F_{\mu\nu} \equiv \overset{\circ}{\nabla}_\mu A_\nu - \overset{\circ}{\nabla}_\nu A_\mu = \partial_\mu A_\nu - \partial_\nu A_\mu$.
No gauge boson couplings with distortions appear in this kinetic term. The same is true for non-Abelian gauge bosons.

A possible interaction between distortion and the Abelian gauge field is the kinetic mixing term. 
Defining $P_{\mu\nu}=\partial_\mu P_\nu-\partial_\nu P_\mu$, one can introduce $\sim P_{\mu\nu} F^{\mu\nu}$ without breaking projective invariance.
As noted in Sec.~\ref{sec:vec}, a kinetic mixing of the form $\sim N_{\mu\nu}F^{\mu\nu}$ (\ref{kinmix}) is not consistent with general projective symmetry, but consistent with the scalar projective symmetry.

%%%%%%%%%%%%%%%%%%%%%%%%%%%%%%%%%%%%%%%
\section{Distortion decay width}
\label{sec:decay}
%%%%%%%%%%%%%%%%%%%%%%%%%%%%%%%%%%%%%%%

In the model of discussed in Sec.~\ref{sec:scalar}, a scalar distortion $\varphi$ is naturally light and becomes dynamical. It mixes with the Higgs boson and decays into two photons.
Assuming that its mass is much lighter than all electrically charged fermions, the decay rate is given by~\cite{Gunion:1989we}
\begin{align}
    \Gamma(\varphi\to 2\gamma) = \theta^2\frac{\alpha_e^2 m^3}{144\pi^3 v^2} \left[\sum_iQ_i^2N_{c,i}-\frac{21}{4}\right]^2,
\end{align}
where $\theta= c_Hvm/m_h^2$ is the mixing angle between $\varphi$ and the Higgs, $\alpha_e=e^2/(4\pi)$ is the fine structure constant, $v=246\,$GeV is the Higgs VEV, $Q_i$ is electric charge of the fermion $i$, $N_{c,i}$ is the color factor of the fermion: $N_{c,i}=1\,(3)$ for leptons (quarks).
The factor $21/4$ corresponds to the contribution from the $W$-boson loop.
For fermion species $i$, we include heavy quarks $c,b,t$ and leptons $e,\mu,\tau$.
We do not include light quarks $u,d,s$ in this summation since their effect is not incorporated in this way due to the strong dynamics. 
However, the resulting inaccuracy of the estimation of the decay rate $\varphi\to 2\gamma$ is expected to be irrelevant for our purpose~\cite{Piazza:2010ye}.

Note that in some of the mass range plotted in Fig.~\ref{fig:cH}, the $\varphi$ decay into neutrino pair is also kinematically allowed. One can deduce the decay rate from the effective Lagrangian of the form $\mathcal L \sim (L H)^2/\Lambda$, where $L$ denotes the lepton doublet and $\Lambda$ is cutoff scale to give a correct value of the neutrino mass.
Then it is not difficult to see that the decay rate into the neutrino pair is suppressed by $\sim \left(10^3 m_\nu^2/(vm)\right)^2$, with $m_\nu$ being the neutrino mass, compared with 2 photon decay.
It is always negligibly small whenever the decay into neutrinos is accessible.

%%%%%%%%%%%%%%%%% Ref %%%%%%%%%%%%%%%%%%%%%%%%
%\bibliographystyle{utphys}
\bibliography{ref}
%%%%%%%%%%%%%%%%%%%%%%%%%%%%%%%%%%%%%%%%%%%%%%

\end{document}